# Establishing process-structure linkages using Generative Adversarial Networks

Mohammad Safiuddin[a], CH Likith Reddy[b], Ganesh Vasantada[c], CHJNS Harsha[d], Srinu Gangolu[e,*]

[a]*National Institute of Technology Calicut*
[b]*National Institute of Technology Calicut*
[c]*National Institute of Technology Calicut*
[d]*National Institute of Technology Calicut*
[e]*National Institute of Technology Calicut*

**Abstract**

The microstructure of material strongly influences its mechanical properties and the microstructure itself is influenced by the processing conditions. Thus, establishing a Process-Structure-Property relationship is a crucial task in material design and is of interest in many engineering applications. We develop a GAN (Generative Adversarial Network) to synthesize microstructures based on given processing conditions. This approach is devoid of feature engineering, needs little domain awareness, and can be applied to a wide variety of material systems. Results show that our GAN model can produce high-fidelity multiphase microstructures which have a good correlation with the given processing conditions.

## 1. Introduction

A significant emphasis in the design of advanced material systems is the delineation of process-structure-property relationships [1]. Although analytical and statistical methods for the design of some materials have been successfully used [2], the underlying assumptions of homogeneity restrict their ability to generalize to other material systems. As solutions to these challenges, machine learning and data-driven approaches have piqued interest in the field of material science. Microstructure reconstruction is important in computational material design because it provides an effective method for understanding the high-dimensional microstructure space. Deep learning has been used in the past to model material properties from microstructures [3] and to synthesise microstructures with desirable properties [4, 5].

*Corresponding author. Tel.: +91 495 2286437
*URL:* srinu@nitc.ac.in (Srinu Gangolu)

Deep learning-based microstructure reconstruction relies heavily on generative models like variational autoencoders (VAE) [6] and generative adversarial networks (GAN) [7]. VAE was used by Cang et al. [4] to create two-phase microstructures, demonstrating the ability of convolutional networks to model material characteristics from microstructures. To synthesise microstructures, Yang et al. [5] used deep convolutional GAN and transfer learning to enhance structure-property predictions.

The aforementioned line of work does not account for two aspects of the problem at hand. Previous research has concentrated on two-phase microstructures, while multiphase microstructures are present in many material systems. Characterization and generation of multiphase microstructures have been studied rarely, owing to the difficulty of determining the higher order correlation of multiphase materials. Also, recent research has not considered the impact of processing conditions on the microstructure. This is a critical component of material design since the interest is to not only design an ideal microstructure but also to figure out how to produce it.

We overcome these two problems by creating a GAN architecture capable of synthesising multiphase alloy microstructures from processing conditions and illustrate this approach by using the Ultra High Carbon Steel Database [8]. This is the first attempt, to our knowledge, to represent the processing-structure relationship as a multi-conditional image synthesis problem.

## 2. Methodology

### *2.1. The Ultra High Carbon Steel Database (UHCSDB)*

The UHCS database contains 961 microstructures produced by scanning electron microscopy (SEM) of samples with similar compositions but different processing conditions. These steels have a carbon content of 1-2.1% per cent, and the steel grade used in this dataset is Adamite grade steel (AS100). Due to the increased carbon concentration of these alloys in comparison to standard steels, pro eutectoid cementite ($Fe_3C$) produces a carbide network coupled with the grain boundaries of the high-temperature austenite process and is a distinguishing microstructure feature of these alloys. These brittle, powerful carbides contribute to the well-known strength and wear resistance of ultra-high carbon steel [9]. Various microconstituents found in the dataset are shown in figure 1.

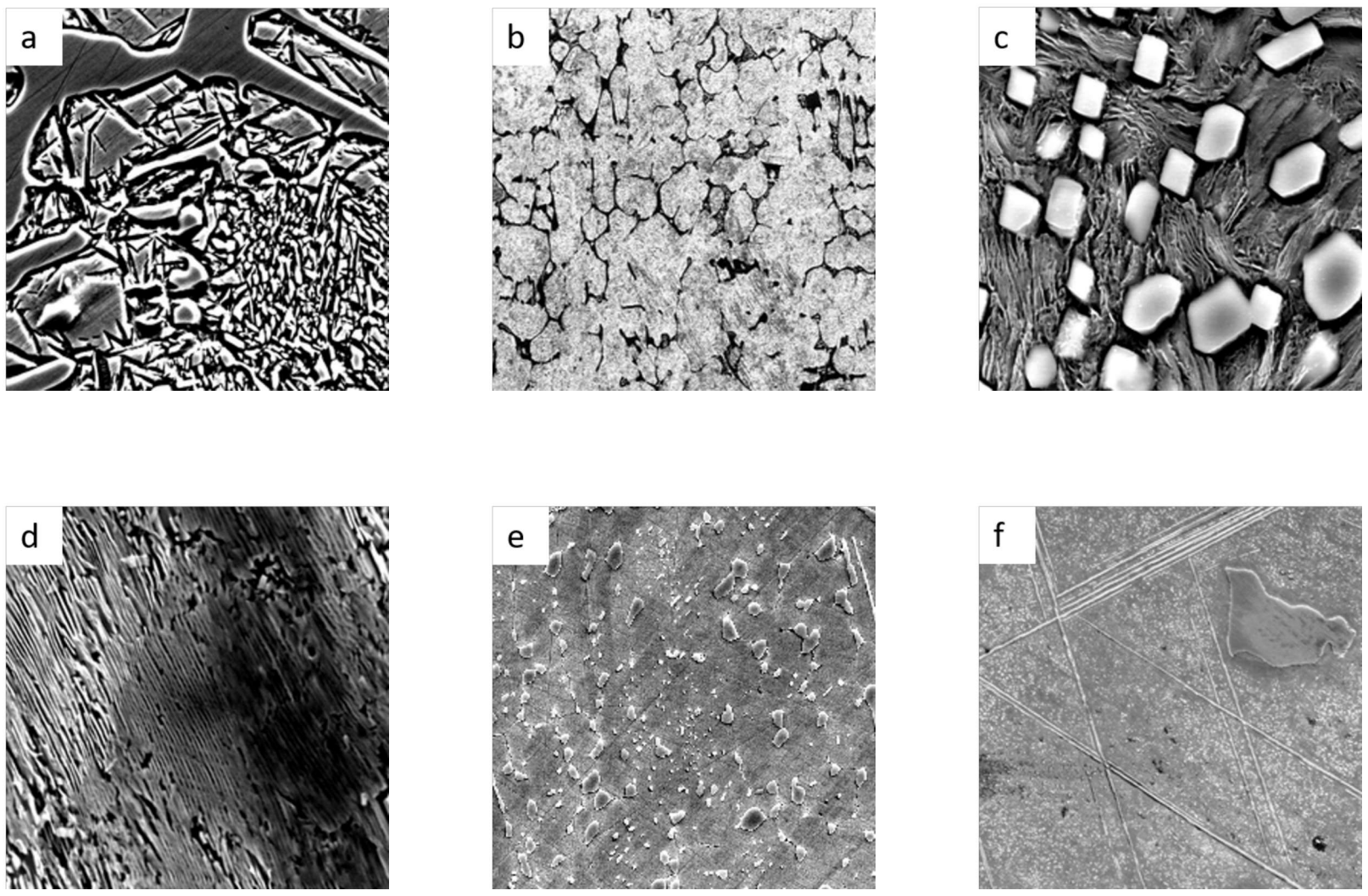

Figure 1: Primary microstructure constituents in the UHCS dataset: a) Pearlite, b) Cementite Network, c) Spheroidite, d) Pearlite + Spheroidite, e) Widmanstätten, f) Martensite

Out of the 961 images present, 598 images have metadata on processing conditions in the form of three parameters *i.e.* annealing temperature, anneal-ing time and the cooling method. It is evident from figure 2 that spheroidite is the major micro constituent in many classes. This imbalance poses an issue to the GAN which can be solved by eliminating samples from the majority class (under-sampling) or adding more examples from the minority class (over-sampling). Over-sampling can cause overfitting while under-sampling can cause a loss in information. Therefore, when sampling from an unbalanced dataset, we rebalance the class distributions and mitigate overfitting using data augmentations. Owing to the small size of the dataset, using a deep learning model like GAN effectively, is a challenging task. We can overcome this to an extent using DiffAugment [10] with Colour, Translation and Cutout policy.

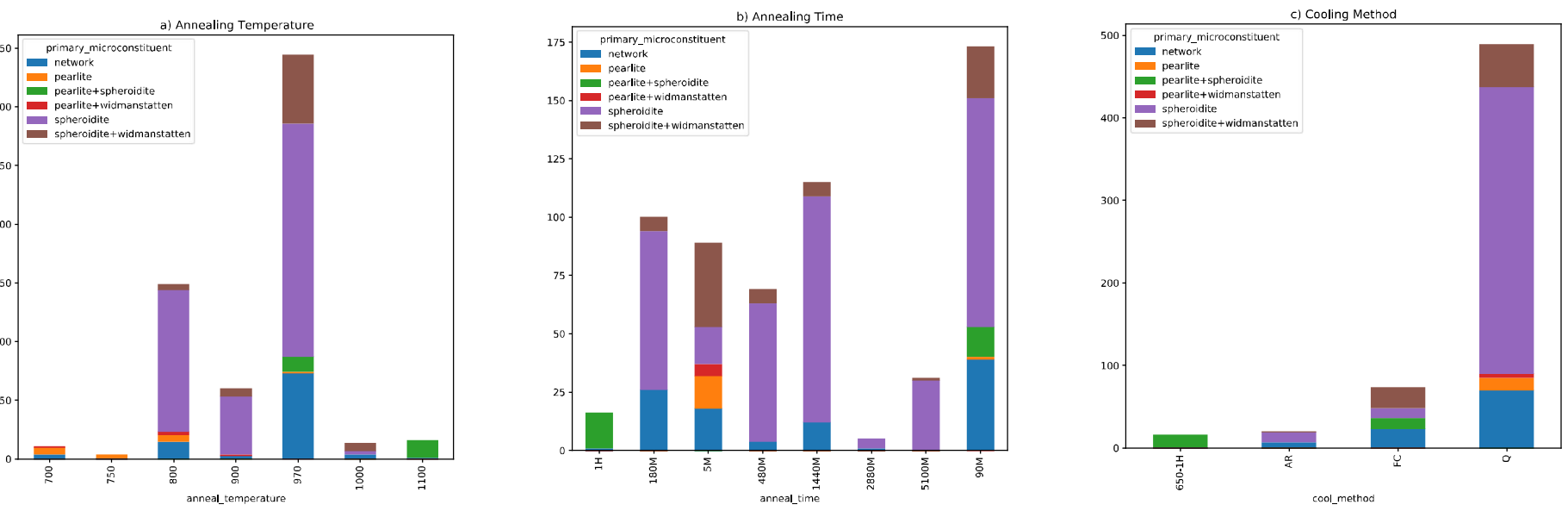

Figure 2: Stacked bar plots for process parameters. Imbalance in the dataset can be clearly seen.

### *2.2. Learning Framework*

To learn underlying latent distributions from high dimensional data, such as images, GANs formulate the learning problem as a two-player zero-sum game between a generator which tries to produce synthetic images indistinguishable from real ones and a discriminator which tries to differentiate if an image is real or has been synthesised. Conditional generative models also offer an effective way to better manage features in synthesised samples by imposing additional structure onto the GAN latent space. Conditional GAN [11] suggests an adaptation of this technique by supplying side details (class labels) to both the generator and discriminator. This increases the consistency and variety of the synthesised images. Following Goodfellow et al's [7] introduction of GAN, several improvements have been proposed in order to promote consistent training and rapid convergence. An overview of the GAN framework is shown in figure 3.

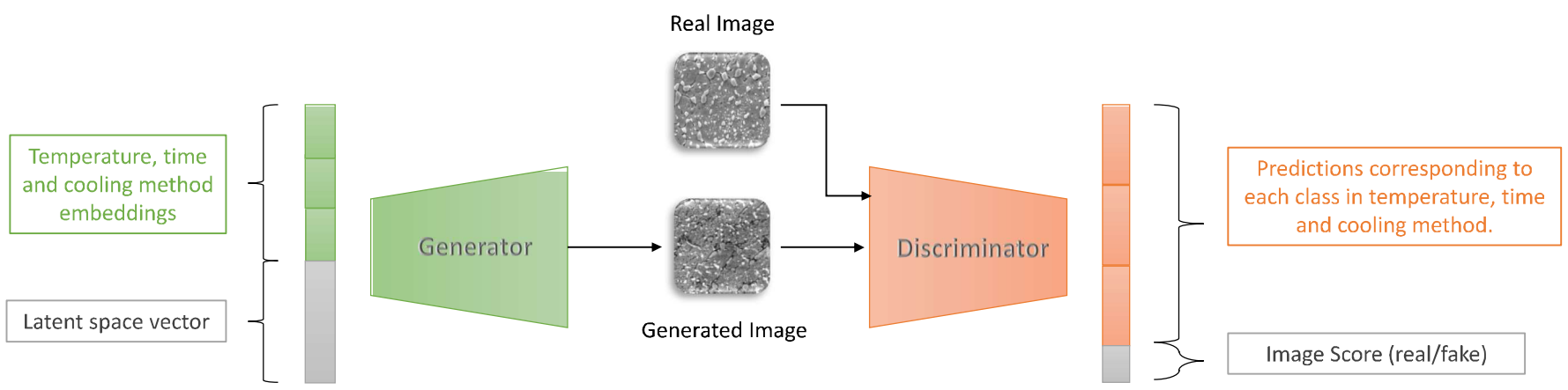

Figure 3: Overview of the GAN framework used to synthesize microstructures based on processing conditions.

*Architecture.* BigGAN [12] improves efficiency by introducing general architectural improvements for better scalability and by modifying a regularisation scheme to improve conditioning. This current work builds on the BigGAN-deep architecture to create microstructures from specified processing parameter values. The generator's input mechanism was modified to accommodate three conditional inputs as opposed to just one. The discriminator was modified to output a 21-dimensional vector (refer to Loss function) as output instead of a scalar. The aforementioned modifications are described in more detail in Appendix A.

*Loss function.* We use Omni loss [13] instead of hinge-loss used in the BigGAN paper [12]. Loss function can be elucidated as follows.

Let $\hat{y}$ and $y$ denote a classifier output and the corresponding multi-label vector respectively. . Assume you have $K$ positive labels and $L$ negative labels. Then $\hat{y}$ is a score vector with $K + L$ dimensions.. The omni-loss is defined as

$$\mathcal{L}_{\text{omni}}\left(\hat{\boldsymbol{y}}, \boldsymbol{y}\right) = \log\left(1 + \sum_{i \in I_{\text{neg}}} e^{\hat{y}_i}\right) + \log\left(1 + \sum_{j \in I_{\text{pos}}} e^{-\hat{y}_j}\right) \tag{1}$$

$I_{\text{neg}}$ contains indexes of negative scores (i.e., $|I_{\text{neg}}| = L$ ), and $I_{\text{pos}}$ contains indexes of positive scores (i.e., $|I_{\text{pos}}| = K$ ). $\hat{y}_k$ represents the element $k$ of vector $\hat{y}$. Using Eq. 1 we can define the loss for both the generator and discriminator.

The discriminator loss is composed of two parts, one for $x_{real}$ (from training set) and another for $x_{fake}$ (from generator). The multi-label vector associated with $x_{real}$ is given by

$$\boldsymbol{y}_{\text{real}} = [\underbrace{-1,\ldots,1_{\text{gt}_{\text{temp}}},\ldots,-1}_{C_{temp}},\underbrace{-1,\ldots,1_{\text{gt}_{\text{time}}},\ldots,-1}_{C_{time}},\underbrace{-1,\ldots,1_{\text{gt}_{\text{cool}}},\ldots,-1}_{C_{cool}},\underbrace{1_{\text{real}},-1}_{2}] \tag{2}$$

$C_{temp}, C_{time}, C_{cool}$ are the number of classes in anneal temperature, anneal time and cooling method respectively. The dimension of $\boldsymbol{y}_{\text{real}}$ is given by $C = C_{temp} + C_{time} + C_{cool} + 2 = 7 + 8 + 4 + 2 = 21$. The value of $1_{\text{gt}_{\text{temp}}}$, $1_{\text{gt}_{\text{time}}}$ and $1_{\text{gt}_{\text{cool}}}$ is 1 if index in the vector is equal to the ground truth label of $x_{real}$, else $-1$. 1 is used to denote that the corresponding score belongs to the positive set and $-1$ is used for the negative set. The multi-label vector for $x_{fake}$ is also a 21 dimensional vector and is given by

$$\boldsymbol{y}_{\text{fake}} = [\underbrace{-1,\ldots,-1,\ldots,-1}_{C_{temp}},\underbrace{-1,\ldots,-1,\ldots,-1}_{C_{time}},\underbrace{-1,\ldots,-1,\ldots,-1}_{C_{cool}},\underbrace{-1,1_{\text{fake}}}_{2}] \tag{3}$$

From Eq. 1, 2, 3, we can define discriminator loss as follows

$$\mathcal{L}_D = E_{\boldsymbol{x}_{\text{real}} \sim p_{\text{d}}}[\mathcal{L}_{\text{omni}}(D(\boldsymbol{x}_{\text{real}}), \boldsymbol{y}_{\text{real}})] + E_{\boldsymbol{x}_{\text{fake}} \sim p_{\text{g}}}[\mathcal{L}_{\text{omni}}(D(\boldsymbol{x}_{\text{fake}}), \boldsymbol{y}_{\text{fake}})] \tag{4}$$

The generator intends to deceive the discriminator into thinking the samples are real. As a result, the multi-label vector is

$$\boldsymbol{y}_{\text{fake}}^{(\text{G})} = [\underbrace{-1,\ldots,1_{\text{G}_{\text{temp}}},\ldots,-1}_{C_{temp}},\underbrace{-1,\ldots,1_{\text{G}_{\text{time}}},\ldots,-1}_{C_{time}},\underbrace{-1,\ldots,1_{\text{G}_{\text{cool}}},\ldots,-1}_{C_{cool}},\underbrace{1_{\text{real}},-1}_{2}] \tag{5}$$

The value of $1_{\text{G}_{\text{temp}}}$, $1_{\text{G}_{\text{time}}}$ and $1_{\text{G}_{\text{cool}}}$ is 1 if index in the vector is equal to the label used by the generator to generate $x_{fake}$, else $-1$. The generator loss

is given by

$$\mathcal{L}_G = E_{\boldsymbol{x}_{\text{fake}} \sim p_g} \left[ \mathcal{L}_{\text{omni}} \left( D(\boldsymbol{x}_{\text{fake}}), \boldsymbol{y}_{\text{fake}}^{(\text{G})} \right) \right] \tag{6}$$

*Training.* We use Adam optimizer [14] with a learning rate of 0.0002, $\beta_1 = 0.9$ and $\beta_2 = 0.999$ for both the generator and discriminator. Additionally, the weight decay of the generator and discriminator is set to 0.001 and 0.0005, respectively. Model is trained for 6000 epochs with a batch size of 96 and a $D : G$ ratio of $4 : 1$. The implementation was done using Pytorch [15] and it was trained on the free GPU available on Kaggle [1].

## 3. Results

After adequate training, the GAN model's generator component generates microstructures that visually mimic the intricacy of the original microstructures with surprising realism. While the GAN-generated microstructures and the original data appear to be extremely comparable, a closer examination reveals that the original microstructures are more variable than the synthetic microstructures. This is consistent with observations demonstrating that GANs produce less variance than the data used to train them [16]; they learn the average structure more quickly than the outliers. Nonetheless, the created microstructures resemble those in the training set. Also, It is evident from figure 4 that there a strong correlation between the processing parameters and the generated microstructures. The characteristics of the generated microstructures that are worth noting are described in Discussion .

*Notation for processing conditions.* Let us consider: 800-5-Q, where the anneal temperature is 800°C, anneal time is 5 minutes and cooling method is Quenching. Cooling methods present in the dataset are Quenching, Air Cooling, Furnace Cooling and constant heating at 650°C for 1 hour which are represented as Q, AR, FC, 650-1H respectively.

[1] `www.kaggle.com`

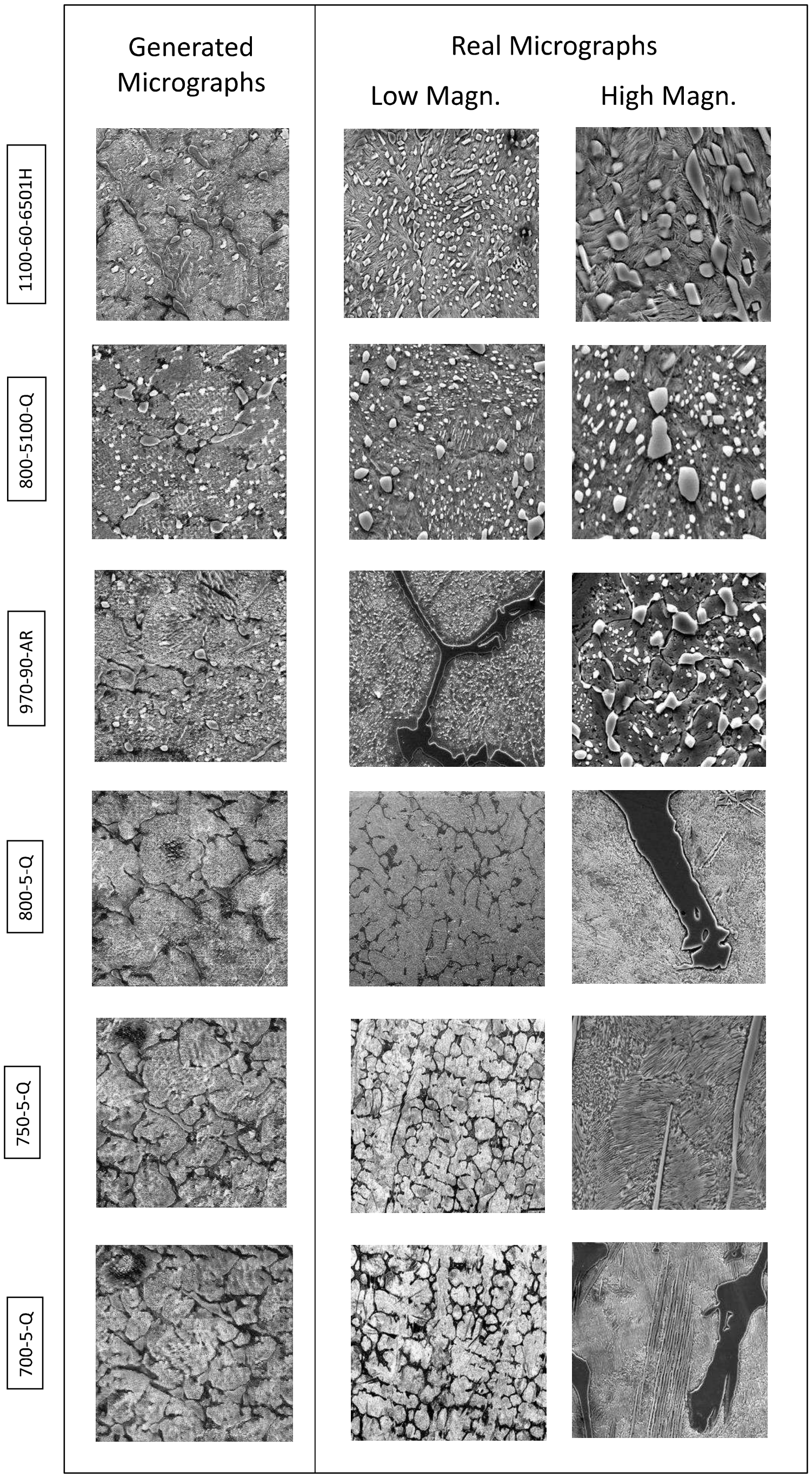

Figure 4: A visual comparision of generated microstructures with the real ones in the dataset.

## 4. Discussion

*Effect of Magnification.* The training data consists of images with varying magnifications. The current GAN model adapts to this in a peculiar manner. The relative scale of a certain feature with respect to other feature in a generated microstructure is similar to the microstructures in the dataset. Let us consider the case represented in figure 5. Spheroidite grains can be clearly seen in figure 5c. However, there is a network structure in figure 5b and the spheroidite is not clearly visible. The generated image has a level of magnification where both spheroidite and network can be seen. The size of spheroidite grains is much smaller compared to the ones in figure 5c and the network web is less spread out compared to figure 5b. This adaptability is seen across other processing parameters as well (see figure 4).

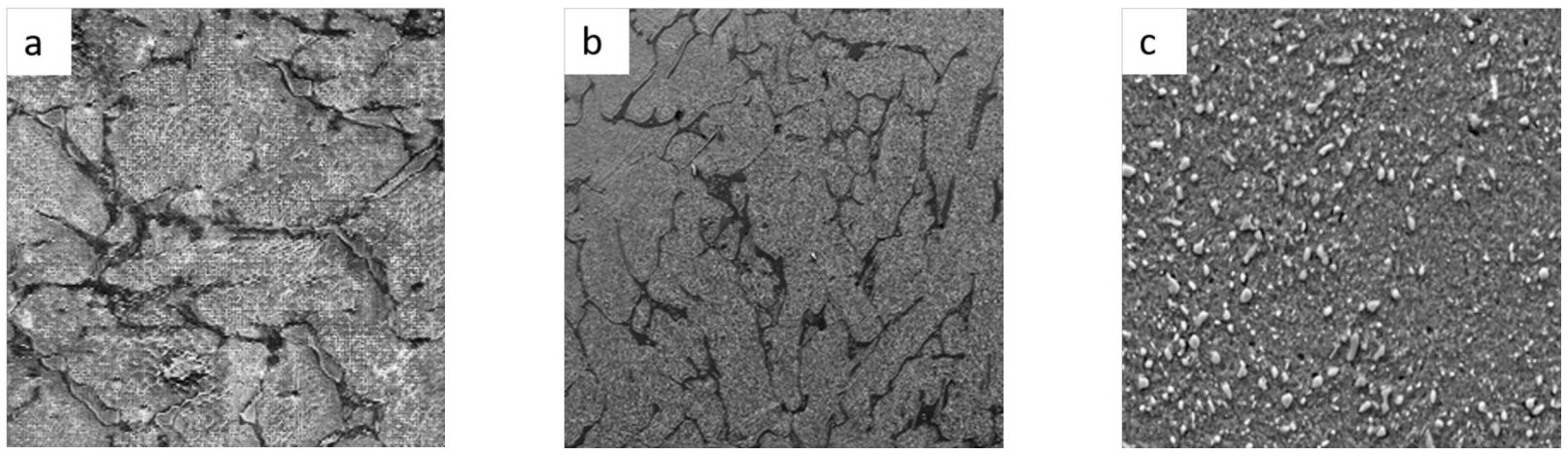

Figure 5: Microstructures corresponding to 800-1440-Q: a) Generated, b) Real (Magn. 98x), c) Real (Magn. 1964x).

*Microstructure Evolution.* To understand if the model has learnt any physically accurate patterns based on the changes in the processing conditions, let us consider the case in figure 6. It can be observed that, as the annealing time increases, the size of the sphere keeps on increasing by combining with nearby spheres. This phenomenon is in line with the work of Hecht et al.[17] though not completely consistent from 60M to 90M.

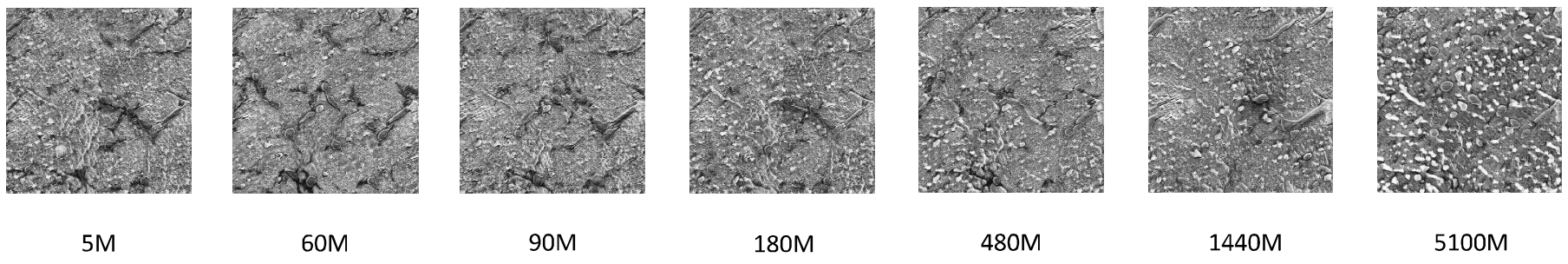

Figure 6: Generated microstructures corresponding to 1000-$X$-Q, where $X$ is increased from 5 minutes to 5100 minutes.

## 5. Conclusion

We used a modified version of the BigGAN-deep architecture with a re-framed Omni loss to generate microstructures under specific conditions. We demonstrate the fidelity of the generated microstructures and their correlation with the processing methods used. The GAN model, in our opinion, offers the following advantages:

- The GAN framework can be used to model any type of material system.
- The generator once trained can generate an arbitrarily large number of unique microstructures.
- Although the GAN model needs a substantial amount of computing power to train, it is quite efficient when it comes to generating synthetic microstructures.

Despite being trained on an extremely small dataset (598 images), our GAN model offers decent fidelity. We believe that the variation and quality of the images can be improved by using a large dataset. Also, the trained model can be used in transfer learning for other steel microstructure datasets.

This GAN framework can be included into a larger framework to establish Process-Structure-Property (PSP) linkages. The generated microstructures can be added to an already available dataset to improve performance for classification of microconstituents [18].

## 6. Acknowledgement

## 7. Data availability

The data used in this work was publicly made available by DeCost et al. [8] here. The pre-trained model and code used in this work can be found here.

## Appendix A. Architectural details

In this work, we aim to establish a processing-structure relationship via multi-conditional image generation. BigGAN is more intuitive for multi-conditional scenario since it provides class conditional information at multiple levels of the generator as opposed to just providing it at the initial level which is the case for typical GAN architectures. This is illustrated in figure A.7.

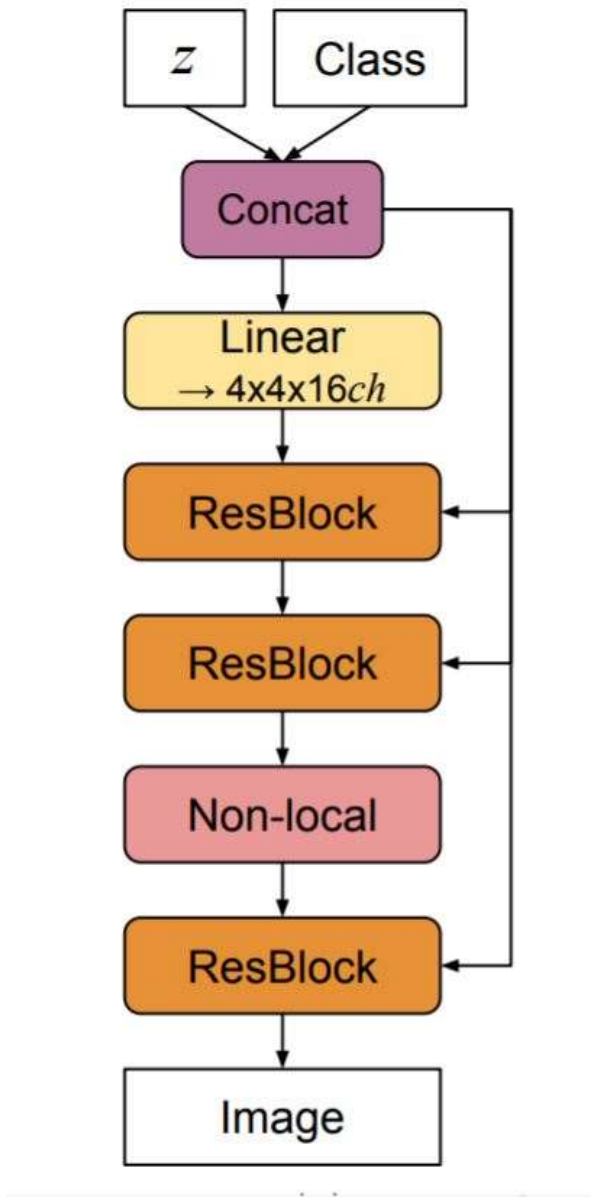

Figure A.7: Layout of the architecture for BigGAN-deep generator [12]

As mentioned earlier, we modify the BigGAN-deep to be better suited for the task at hand. We use three embeddings for annealing temperature, annealing time and cooling method as opposed to just one. We are not using a projection style generator . We observed that the model collapsed fairly quickly with a default latent dimension of 128. Therefore, we increase the latent space dimension to 384 to avoid collapse as increasing the latent space dimension gives the model more degrees of freedom to fit the individual samples, thereby increasing the quality of generated images. The aforementioned modifications can be observed by comparing figure A.8 and A.9.

| |
|---|
| $z \in R^{384} \sim \mathcal{N}(0, I)$<br>Embed $(y_{temp}) \in R^{128}$<br>Embed $(y_{time}) \in R^{128}$<br>Embed $(y_{cool}) \in R^{128}$ |
| Linear $(384 + (128 \times 3)) \rightarrow 4 \times 4 \times 16ch$ |
| ResBlock $16ch \rightarrow 16ch$ |
| ResBlock up $16ch \rightarrow 16ch$ |
| ResBlock $16ch \rightarrow 16ch$ |
| ResBlock up $16ch \rightarrow 8ch$ |
| ResBlock $8ch \rightarrow 8ch$ |
| ResBlock up $8ch \rightarrow 8ch$ |
| ResBlock 8ch $\rightarrow 8ch$ |
| ResBlock up $8ch \rightarrow 4ch$ |
| Non-Local Block $(64 \times 64)$ |
| ResBlock $4ch \rightarrow 4ch$ |
| ResBlock up $4ch \rightarrow 2ch$ |
| ResBlock $2ch \rightarrow 2ch$ |
| ResBlock up $2ch \rightarrow ch$ |
| BN, ReLU, $3 \times 3$ Conv $ch \rightarrow 1$ |
| Tanh |

**Generator**

| |
|---|
| Image $x \in R^{256 \times 256 \times 1}$ |
| $3 \times 3$ Conv $1 \rightarrow ch$ |
| ResBlock down $ch \rightarrow 2ch$ |
| ResBlock $2ch \rightarrow 2ch$ |
| ResBlock down $2ch \rightarrow 4ch$ |
| ResBlock $4ch \rightarrow 4ch$ |
| Non-Local Block $(64 \times 64)$ |
| ResBlock down $4ch \rightarrow 8ch$ |
| ResBlock $8ch \rightarrow 8ch$ |
| ResBlock down $8ch \rightarrow 8ch$ |
| ResBlock $8ch \rightarrow 8ch$ |
| ResBlock down $8ch \rightarrow 16ch$ |
| ResBlock $16ch \rightarrow 16ch$ |
| ResBlock down $16ch \rightarrow 16ch$ |
| ResBlock $16ch \rightarrow 16ch$ |
| ReLU, Global sum pooling |
| $linear \rightarrow 21$ |

**Discriminator**

Figure A.8: Our Architecture for 256 × 256 images. $ch = 64$

| Generator |
|---|
| $z \in R^{128} \sim \mathcal{N}(0, I)$<br>Embed $(y) \in R^{128}$ |
| Linear $(128 + 128) \to 4 \times 4 \times 16ch$ |
| ResBlock $16ch \to 16ch$ |
| ResBlock up $16ch \to 16ch$ |
| ResBlock $16ch \to 16ch$ |
| ResBlock up $16ch \to 8ch$ |
| ResBlock $8ch \to 8ch$ |
| ResBlock up $8ch \to 8ch$ |
| ResBlock 8ch $\to 8ch$ |
| ResBlock up $8ch \to 4ch$ |
| Non-Local Block $(64 \times 64)$ |
| ResBlock $4ch \to 4ch$ |
| ResBlock up $4ch \to 2ch$ |
| ResBlock $2ch \to 2ch$ |
| ResBlock up $2ch \to ch$ |
| BN, ReLU, $3 \times 3$ Conv $ch \to 3$ |
| Tanh |

| Discriminator |
|---|
| Image $x \in R^{256 \times 256 \times 3}$ |
| $3 \times 3$ Conv $1 \to ch$ |
| ResBlock down $ch \to 2ch$ |
| ResBlock $2ch \to 2ch$ |
| ResBlock down $2ch \to 4ch$ |
| ResBlock $4ch \to 4ch$ |
| Non-Local Block $(64 \times 64)$ |
| ResBlock down $4ch \to 8ch$ |
| ResBlock $8ch \to 8ch$ |
| ResBlock down $8ch \to 8ch$ |
| ResBlock $8ch \to 8ch$ |
| ResBlock down $8ch \to 16ch$ |
| ResBlock $16ch \to 16ch$ |
| ResBlock down $16ch \to 16ch$ |
| ResBlock $16ch \to 16ch$ |
| ReLU, Global sum pooling |
| Embed$(y) \cdot h +$<br>$(linear \to 1)$ |

Figure A.9: Original BigGAN-deep architecture for $256 \times 256$ images. $ch = 64$